\documentclass[a4paper]{jpconf}
\usepackage{graphicx}
\newcommand{\sfig}[2]{
\begin{figure}[h]
\includegraphics[width=18pc]{#1}\hspace{2pc}%
\begin{minipage}[b]{18pc}\caption{\label{fig:#1}#2}
\end{minipage}
\end{figure}
}

\usepackage{fancyhdr}
\begin{document}

\title{Dark energy physics expectations at DES}
\author{Marcelle Soares-Santos \\ (for the DES Collaboration)}
\address{Center for Particle Astrophysics, 
         Fermi National Accelerator Laboratory, Batavia IL 60510}
\ead{marcelle@fnal.gov}

\begin{abstract}
Giving rise to a new and exciting research field,
observations of the last 13 years established the accelerated
expansion of the Universe. This is a strong indication of new physics, 
either in the form of a new energy component of the Universe 
-- dark energy -- or of theories of gravity beyond general relativity.
A powerful approach to this problem
is the study of complementary cosmological probes in large optical galaxy surveys
such as the Dark Energy Survey (DES).
We present the expectations for dark energy physics based on the combination of 
four fundamental probes:
galaxy clusters, weak lensing, large scale structure and supernovae. 
We show that DES data have constraining power to improve current measurements of the dark 
energy equation-of-state parameter by a factor of 3--5 and to distinguish between 
general relativity and modified gravity scenarios.
\end{abstract}

\thispagestyle{fancy}

\section{Introduction}\label{intro}
The accelerated expansion of the Universe is a well established fact \cite{Frieman:2008}, 
but although the dark energy density $\Omega_{\Lambda}$ has been determined to a few 
percent precision, getting at its nature is more challenging. This requires
measurements of its equation-of-state, e.g.~in terms of the parameters $w_0$ and
$w_a$ of the phenomenological model 
$w(a) = w_0 + w_a (1-a)$, 
and tests of general relativity (GR), e.g.~through the 
$\gamma$ parameter \cite{Linder:2005}.
Currently $w_0$ is known to 15\% while $w_a$ is largely 
unconstrained \cite{Sullivan:2011} and measurements of $\gamma$  
can not yet distinguish between GR and modified gravity \cite{Thomas:2009,Reyes:2010}.
The Dark Energy Survey 
(DES, {\tt{darkenergysurvey.org}})
\cite{DES:2005} is a ground-based photometric survey conceived to 
significantly improve such measurements by combining \cite{Albrecht:2006} 
galaxy clusters, weak lensing (WL), large scale structure (LSS) and 
Type Ia supernovae (SNe).  

The DES collaboration has 
built the Dark Energy Camera (DECam) \cite{Flaugher:2010}, 
an imaging instrument comprised of 74 250 micron thick CCDs \cite{Estrada:2010}
to be installed on the Cerro Tololo Inter-American Observatory (CTIO) 4-meter 
Blanco telescope \cite{Abbott:2006}. 
DES will use  DECam for 525 nights in the 2012--2017 austral spring/summer 
to survey a 5000 deg$^2$ area of the sky in 5 filters $grizY$ up to redshift 
$z\simeq 1.5$ achieving a volume of 24 $h^{-3}$Gpc$^{3}$, 7 times larger 
than the largest existing CCD survey of the Universe by volume to date 
\cite{Thomas:2011}. DES data  include:
photometric redshifts and shapes of 300 million galaxies, 
mass and spatial distribution of 100,000  
clusters and detection of 4000 Type Ia SNe.

This paper presents the dark energy science prospects for this data set. 
Section \ref{probes} discusses each probe.
Section \ref{combined} explores the of combination these observables to 
determine as well $w_a$ and distinguish between GR and modified gravity. 
Planck priors and statistical errors only  are assumed throughout the paper. 
Conclusions are drawn in Section \ref{conclusions}.

\section{Probes of cosmic acceleration}\label{probes}

\subsection{Weak gravitational lensing}
Images of distant galaxies show distortions (shear) due to the gravitational bending of 
their light by structures along the line-of-sight. This weak lensing  
effect allows us to measure the mass of foreground structures (e.g.~galaxy clusters)
using shear radial profiles \cite{Johnston:2007p,Simet:2011}.
But the overall cosmic shear field can also be measured, 
by dividing the survey area in pixels and averaging the galaxy shapes in each pixel. 
The statistical signal  from the cosmic shear field
(shear-shear correlation function or, in Fourier space, the power spectrum)
allows us to determine \cite{Fu:2008,Lin:2011} 
the total matter content 
of the Universe $\Omega_m$ and
the normalization of matter fluctuations $\sigma_8$. 
With similar depth and an area 18 times larger than the largest survey used for 
cosmic shear 
measurements to date \cite{Lin:2011,Annis:2011}, DES will be sensitive to the effect of 
dark energy on the cosmic shear.

\subsection{Galaxy clusters}
The number density of the largest gravitationally bound structures (clusters) as a 
function of redshift and mass
is sensitive to the cosmic expansion history because the 
comoving volume element (geometry) of the Universe is changing and because 
initially small density perturbations are evolving to form them 
through gravity against that expansion (growth of structure). Geometry is 
dominant for clusters at $z<0.6$ while growth of structures dominates at higher
redshifts \cite{Frieman:2008}. 
Clusters up to $z \simeq 1.0$ can be efficiently detected 
in DES data \cite{Soares-Santos:2010}  as enhancements in 
the surface density of galaxies using  photometric redshifts estimated from 
their colors to substantially reduce projection effects. The cluster richness, 
defined as the number of member galaxies, is a good proxy for its mass and the   
mass-richness relation can be calibrated using weak lensing 
\cite{Johnston:2007p,Simet:2011}. 
Optically selected clusters 
have been used to 
measure $\Omega_m$ and $\sigma_8$ \cite{Rozo:2010}. By obtaining a complete and 
pure sample of galaxy clusters up to redshift $\simeq 1$ over 5000 square 
degrees, DES allows us to extend these measurements to 
constrain dark energy models. 

\subsection{Large scale structure}
Gravity-driven acoustic oscillations of the coupled 
photon-baryon fluid in the early Universe leave an imprint on the mass power 
spectrum that can be detected as an excess in the galaxy-galaxy 
correlation function at a characteristic scale. 
The scale of the oscillations is that of the sound horizon at the epoch of recombination,
known through the power spectrum of cosmic microwave background 
anisotropies  measured by WMAP \cite{Jarosik:2011} to be $146.2 \pm 1.1$ Mpc.
The corresponding scale on the galaxy angular power spectrum is a geometric 
probe of the cosmic expansion history and has been used to measure dark energy
\cite{Percival:2010,Blake:2011}.  
DES will use LSS to perform such measurements from its data set.  

\subsection{Supernovae}
Type Ia SNe can be used as standard candles to probe the cosmic acceleration 
through the magnitude-redshift relation (the Hubble diagram). Their 
progenitors,
carbon-oxygen white dwarfs accreting mass from a companion star, explode 
producing $^{56}$Ni (which later decays into $^{56}$Co) at a rate 
directly related to their peak luminosity \cite{Frieman:2008}.  
The first direct evidence for cosmic acceleration came from SNe Ia, and of the four
probes, they have 
provided the strongest constraints on the dark energy equation-of-state parameter 
to date \cite{Sullivan:2011}.  
By observing about 10 times more SNe, DES will significantly improve such 
measurements \cite{Bernstein:2011}.  

\section{Combined constraints}\label{combined}
The  cosmological techniques explored by DES have constraining power 
to probe dark energy with high precision individually, but it is the combination 
of these complementary probes that can produce the best results. Such combinations
have been explored to some extent (e.g., SNe+CMB+LSS \cite{Sullivan:2011})
but DES is the first experiment to combine all four probes from the same data set,
being able to achieve percent-level uncertainty on $w_0$ and, in addition,
measure $w_a$. 
By combining the four probes we can measure $w_0$ at 5\% and $w_a$ at 30\%
uncertainty level, as shown in Fig.~\ref{fig:desforecast.pdf},
improving the constraints on the dark energy 
equation-of-state $w(a)$ by a factor of 3-5 with respect to current experiments.

\sfig{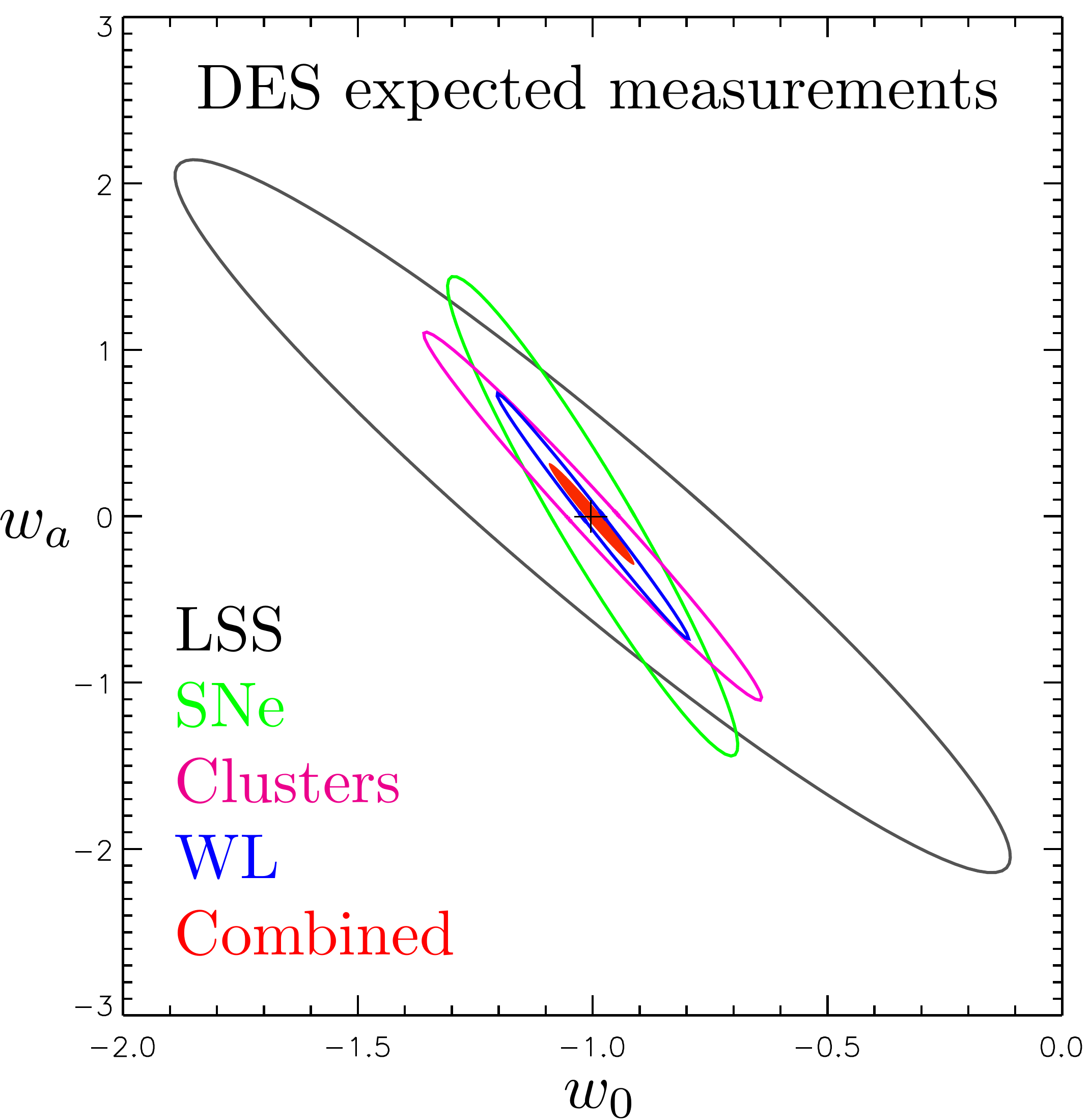}{Forecast for 1$\sigma$ constraints on dark energy parameters 
from the DES probes, including only statistical errors
and assuming $w_0=-1$, $w_a=0$ as the true model \cite{DES:2006,Shapiro:2010}. 
 Each individual constraint uses Planck priors. 
The supernovae constraint includes an 8\% prior on $H_0$. The constraints from
the combination of the four probes (solid red region) correspond to 
uncertainties in $w_0$ and $w_a$ of 5\% and 30\% respectively.}

Our data set also allows us to distinguish between GR and 
certain modified gravity theories, by measuring the parameter $\gamma$. 
This can be achieved using a multi-dimensional
consistency test of the four dark energy probes \cite{Shapiro:2010}. 
An inconsistency would result in
contours slightly miscentered with respect to each other.   
Such analysis, performed on DES data, can distinguish between $\gamma = 0.55$ (GR case) 
and
$\gamma=0.68$ (approximately the value for the Dvali-Gabadadze-Porrati (DGP) 
braneworld model \cite{DGP:2000}) at a 99.1\% level \cite{Shapiro:2010}.  

\section{Conclusions}\label{conclusions}
DES is a photometric survey designed to shed light on the dark energy problem 
through four complementary methods (LSS, SNe, Clusters and Weak Lensing).
Commissioning of the  DES imaging instrument, DECam, is imminent. The survey is scheduled 
to start in the second semester of 2012, take data over 5 years and make available to
the astronomical community a data set of unprecedented depth for its area 
(5000 deg$^2$ up to redshift $\simeq$ 1.5). 
This rich data set has the potential for a variety of studies, from 
galaxy evolution to cosmology.
The prospects for dark energy science are highlighted in this paper with focus on 
the key analyses of the four cosmological probes to  
improve current measurements of the equation-of-state parameter $w(a)$ by 
a factor of 3-5. 
DES also has the potential to distinguish between GR and 
modified gravity theories by measuring, for instance, deviations of the parameter 
$\gamma$ from the GR value $\gamma =0. 55$ at high significance level.
  
\ack Funding for the DES Projects has been provided by the U.S. Department of Energy, the U.S. National Science 
Foundation, the Ministry of Science and Education of Spain, the Science and Technology Facilities Council of the 
United Kingdom, the Higher Education Funding Council for England, the National Center for Supercomputing 
Applications at the University of Illinois at Urbana-Champaign, the Kavli Institute of Cosmological Physics at the 
University of Chicago, Financiadora de Estudos e Projetos, Funda{\c{c}}\~{a}o Carlos Chagas Filho de Amparo 
\`a Pesquisa do Estado do Rio de Janeiro, Conselho Nacional de Desenvolvimento Cient{\'{\i}}fico e 
Tecnol{\'{o}}gico and the Minist\'erio da Ci\^encia e Tecnologia, the Deutsche Forschungsgemeinschaft and the 
Collaborating Institutions in the Dark Energy Survey.

The Collaborating Institutions are Argonne National Laboratories, the University of California at Santa Cruz, 
the University of Cambridge, Centro de Investigaciones Energeticas, Medioambientales y Tecnologicas-Madrid, 
the University of Chicago, University College London, DES-Brazil, Fermilab, the University of Edinburgh, the 
University of Illinois at Urbana-Champaign, the Institut de Ciencies de l'Espai (IEEC/CSIC), the Institut de 
Fisica d'Altes Energies, the Lawrence Berkeley National Laboratory, the Ludwig-Maximilians Universit\"at and the 
associated Excellence Cluster Universe, the University of Michigan, the National Optical Astronomy Observatory,
the University of Nottingham, the Ohio State University, the University of Pennsylvania, the University of 
Portsmouth, SLAC, Stanford University, the University of Sussex, and Texas A \& M University.

\section*{References}
\bibliography{bib2,bib,apj-jour}

\end{document}